\documentclass[numbered,titlepage,oneside,12pt]{article}
\usepackage[paper=a4paper,top=1.0in,bottom=1.0in,right=0.2in,left=0.5in]{geometry}

\usepackage[tiny,rm]{titlesec}
\newpagestyle{trbstyle}{
\sethead{Comert, Rahman, Islam, and Chowdhury}{}{\thepage}
}
\usepackage{lipsum}

\titleformat{\section}{\bfseries}{\thesection{. }}{0pt}{\uppercase}
\titlespacing*{\section}{0pt}{12pt}{*0}
\titleformat{\subsection}{\bfseries}{}{0pt}{}
\titlespacing*{\subsection}{0pt}{12pt}{*0}
\titleformat{\subsubsection}{\itshape}{}{0pt}{}
\titlespacing*{\subsubsection}{0pt}{12pt}{*0}

\usepackage{enumitem}
\setlist[1]{labelindent=0.1in,leftmargin=*}
\setlist[2]{labelindent=0in,leftmargin=*}

\usepackage{ccaption}
\usepackage{amsmath}
\usepackage{amsbsy}
\usepackage{dcolumn}
\usepackage{multirow}
\usepackage[table]{xcolor}
\usepackage{wrapfig}
\usepackage{pifont}
\usepackage{float}
\usepackage{adjustbox}
\usepackage{wrapfig}
\usepackage{soul}
\usepackage{footnote}

\usepackage{enumitem}

\usepackage[switch]{lineno}
\pagewiselinenumbers

\usepackage{graphics}

\usepackage{epstopdf}

\usepackage{caption}
\usepackage{subcaption}

\usepackage{amssymb,amstext,amsmath}
\usepackage{float}
\usepackage{booktabs}
\usepackage{array,ragged2e}
\usepackage[table]{xcolor}
\usepackage{multirow}
\usepackage{algorithm}
\usepackage{algpseudocode}
\usepackage{pifont}

\makeatletter
\renewcommand{\fnum@figure}{\textbf{FIGURE~\thefigure} }
\renewcommand{\fnum@table}{\textbf{TABLE~\thetable} }
\makeatother
\captiontitlefont{\bfseries \boldmath}
\captiondelim{\;}

\usepackage{times}

\usepackage[T1]{fontenc}
\usepackage{textcomp}

\usepackage[sort,numbers]{natbib}

\setcitestyle{round}

\usepackage{graphicx}

\usepackage{atbegshi}
\AtBeginDocument{\AtBeginShipoutNext{\AtBeginShipoutDiscard}}

\usepackage{etoolbox}
\patchcmd{\maketitle}{}{}{}{}

\begin{document}

\begin{nolinenumbers}
\newgeometry{top=1.0in,bottom=1.0in,right=0.5in,left=0.5in}
\title{\textbf{Change Point Models for Real-time V2I Cyber Attack Detection in a Connected Vehicle Environment}}
\author{
\textbf{Paper Number:19-03063}
\\
Gurcan Comert, Computer Science, Physics, and Engineering, Benedict College\\
1600 Harden St., Columbia, SC 29206, Tel:(803) 705 4803, Email:gurcan.comert@benedict.edu\\
\\
Mizanur Rahman, Ph.D.Postdoctoral Fellow,Center for Connected Multimodal Mobility (C2M2)\\
Glenn Department of Civil Engineering, Clemson University\\
127 Lowry Hall, Clemson, SC 29634,Tel: (864) 650-2926; Email: mdr@clemson.edu\\
\\
Mhafuzul Islam, Glenn Department of Civil Engineering, Clemson University\\
351 Flour Daniel, Clemson, SC 29634, Tel: (864) 986-5446\\
Email: mdmhafi@clemson.edu\\
\\
Mashrur Chowdhury, Eugene Douglas Mays Endowed Professor of Transportation\\
Professor of Automotive Engineering, Clemson University\\
Glenn Department of Civil Engineering\\
216 Lowry Hall, Clemson, South Carolina 29634, Tel: (864) 656-3313 \\
Email: mac@clemson.edu\\
Submitted for presentation and publication to \\
the 98th Annual Meeting of the Transportation Research Board\\
January 13-17, 2019\\}

\thanks{\textbf{Acknowledgments:}This material is based on the study supported by the Center for Connected Multimodal Mobility ($C^2M^2$) (USDOT Tier 1 University Transportation Center) Grant headquartered at Clemson University, Clemson, South Carolina, USA. Any opinions, findings, and conclusions or recommendations expressed in this material are those of the author(s) and do not necessarily reflect the views of the Center for Connected Multimodal Mobility, and the U.S. Government assumes no liability for the contents or use thereof. It is also partially supported by U.S. Department of Homeland Security Summer Research Team Program Follow-On grant and NSF Grant No. 1719501.}

\maketitle
\restoregeometry

\end{nolinenumbers}

\newpage
\begin{nolinenumbers}
\section{Introduction}
\label{intro}

From recent reports of National Highway Traffic Safety Administration \cite{NHTSA2013,NHTSA2016}, several benefits are foreseen with connected and autonomous vehicle (CAV) technologies such as up to $80\%$ reduction in fatalities from multi-vehicle crashes and preventing the majority of human error related incidents. In such CAV systems, massive amounts of data will be produced and exchanged between different components through different data communication medium, such as Dedicated Short Range Communication (DSRC), WiFi, 5G, and Long Term Evolution (LTE)~\cite{burt2014big,cvria}. These data can be processed in a cloud, or in an edge computing device at the roadside (i.e., roadside transportation infrastructure) based on different CAV application requirements~\cite{cvria,whaiduzzaman2014survey}. Communication technologies supporting data exchange must also be secured to support CAV operations with specific requirements (e.g., delay, bandwidth, and communication range). With the increase of connectivity in transportation networks, this CAV systems is cognizant of potential cyberattacks~\cite{raya2007securing,USGAO}. 

As cyberattacks are dynamic, it is a challenge to detect security threats in real-time and develop effective countermeasures for connected transportation system \cite{nicolppt}. To increase security and resiliency due to possible attacks or benign system errors, research is needed to investigate detection techniques for different attack types, such as denial of service (DoS), impersonation, false information~\cite{pathre2013novel,mejri2014survey}. Anomaly detection techniques are well-studied in various areas. Specifically, the cybersecurity of firmware updates, cybersecurity on heavy vehicles, vehicle-to-vehicle (V2V) communication interfaces, and trusted vehicle-to-everything (V2X) communications \cite{petit2015potential}. Different type of anomaly detection models exist in literature where recent survey studies related to anomaly detection summarize a comprehensive review of machine learning and rule (signature)-based methods, and their applications to intrusion detection systems (IDS) ~\cite{buczak2016survey,patcha2007overview}. Rule-based attack detection models, originated from cryptography, are abundant especially for their efficiency and computationally light-weight ~\cite{sedjelmaci2014efficient}. However, rule-based models require a detailed understanding of the data generation process and adaptivity or customization based on their respective environment to develop the model. On the other hand, both machine learning and data mining-based attack detection models are adaptable to different attack types both known and unknown patterns. Major concerns here are computational complexity for real-time application, training the model with different cyber attack scenarios, unavailability of cyberattack types in the transportation domain, and determination of update or retraining window. To address these problems, statistical models, specially the change point models, are applicable because of the following advantages: (1) do not require fitting or training; (2) adaptive to different attack data (do not use rules); (3) perform with low data sample sizes; and (4) computationally efficient for real-time applications. Thus, the objective of this study is to investigate the efficacy of two change point models, Expectation Maximization (EM) and Cumulative Sum (CUSUM), for real-time vehicle to infrastructure (V2I) cyberattack detection in a connected vehicle (CV) Environment. To prove the efficacy of these models, we implemented three different type of cyberattacks (i.e., denial of service (DoS), impersonation, and false information)~\cite{van2016survey}, using BSMs generated from CVs through simulation.

\section{Methodology}
\label{sctmethod}
In this study, we investigate statistical change point models, Expectation Maximization (EM) and Cumulative Summation (CUSUM), to detect cyberattacks in a V2I environment.
\subsection{Expectation Maximization Algorithm}
\label{sctEM}
The Expectation Maximization (EM) algorithm is often used to estimate the parameters of mixture models or models with latent variables~\cite{Dempster77maximumlikelihood,hastie2001elements}. Here, EM algorithm is utilized for detecting changes in the process, i.e., cyberattacks~\cite{buczak2016survey}. In sequential framework, the model consists of a hidden state variable (i.e., normal, abnormal) $X_t$ and an observation variable $Y_t$ at time $t$. The state of the model at time $t$ depends only on the previous state of the model at immediate past $X_{t-1}$ through probability distribution $P(X_{t}|X_{t-1})$. Given $N$ data points that are assumed to be generated by mixture of two distributions (i.e., normal and abnormal messages per vehicle per second (MVS), messages per vehicle (MVT), and distance), the EM algorithm is applied to determine the distribution parameters and probability of an cyberattack. 
\subsection{CUSUM Algorithm}
\label{sctCUSUM}
The CUSUM algorithm is commonly used for quality control purposes to detect possible shifts in data generating process characteristics. In cyberattack setting, changes within expected level of deduced measures (MVS, MVT, and distance) are targeted. This paper uses tabular or algorithmic version of the CUSUM rather than control chart. To detect both positive and negative shifts, the two-sided version of the CUSUM algorithm was used. The algorithm works by accumulating positive and negative deviations from a certain target mean, which is commonly taken to be zero. The positive deviations (values above the target) are indicated with $C_{t}^{+}$, and those that are below the target are indicated with $C_{t}^{-}$. The statistics $C_{t}^{+}$ and $C_{t}^{-}$ are referred to as one-sided upper and lower CUSUMs, respectively~\cite{montgomery2009introduction}. The algorithm is then applied to the change point detection of the time series within basic safety messages (BSMs). 
\section{Findings}
\label{sctne}
\subsection{Data Generation for V2I Cyber-attacks}
\label{sctsimulation}
In order to generate the realistic traffic behavior, a microscopic traffic simulation software, Simulation of Urban Mobility (SUMO) is used~\cite{krajzewicz2007simulation}. All vehicle movements are recorded in trace files. A trace file contains time stamp, vehicle ID, latitude, longitude and speed of each vehicle moving on the Perimeter Road, Clemson, South Carolina, which is a part of the Clemson University-Connected and Autonomous Vehicle Testbed (CU-CAVT). The simulation is comprised of $200$ vehicles per hour per lane on Perimeter Road, a four-lane arterial roadway (two lanes each direction) with $35$ miles per hour (mph) speed limit.

Using the generated trace files, three different cyberattack scenarios are generated: (i) denial of service (DoS) attack; (ii) false information attack; and (iii) impersonation attack. For generating DoS attack data, vehicle with ID $6$ is passed a data flooding at 1000 packets per second while other vehicles are sharing data at 10 packets per second to mimic the real-world CV environment where each CV is broadcasts BSMs every one tenth of a second. The total simulation time is $200$ seconds (s). For fake (or false) information attack, false GPS location information (i.e., longitude and latitude) of vehicle ID $2$ are generated simply using Python random variable generation function. To emulate the data for impersonation attack, a false ID for vehicle number $3$ is used as vehicle ID $2$. Two different GPS location and speed information for the vehicle ID $2$ are simultaneously generated. 



Examples of generated attack data are given in Table~\ref{data_tab}. Evident from the table, multilevel attack monitoring could be designed such as vehicle ID by vehicle ID and time with micro level tracking ($0.1$ sec) of data values. The study tracks aggregate measures, such as average message frequency per vehicle per second ($MVS$), average message frequency per vehicle per time interval ($MVT$), distances, and/or track of time series-vehicle speeds and detects changes. 
\begin{table}[H]
\centering
\caption{Examples of attack data generated on RSE}
\label{data_tab}
\begin{tabular}{lrrrrrrrr}
 Type & TS(s) & ID & Lat. & Long. & Speed(m/s) & Pos.(m) & MsgRate \\ 
  \hline
 DoS& 5.10 &   1 & -82.85 & 34.68 & 9.94 & 0.08 & 10.00 \\ 
     & 5.10 &   2 & -82.85 & 34.68 & 8.22 & 0.52 & 10.00  \\ 
     & 5.10 &   3 & -82.85 & 34.68 & 6.21 & 0.74 & 10.00  \\ 
     & 5.10 &   5 & -82.84 & 34.68 & 2.32 & 0.14 & 10.00  \\ 
     & 5.10 &   6 & -82.84 & 34.68 & 0.00 & 0.00 & 10.12  \\ 
     & 5.10 &   6 & -82.84 & 34.68 & 0.00 & 0.00 & 10.23  \\ 
     & 5.10 &   6 & -82.84 & 34.68 & 0.00 & 0.00 & 10.35  \\ 
    \hline
    IMPER.&  1.30 &   1 & -82.85 & 34.68 & 2.65 & 0.00 & 1.00  \\ 
    &   1.30 &   2 & -82.85 & 34.68 & 0.51 & 0.00 & 1.00  \\ 
    &   1.40 &   1 & -82.85 & 34.68 & 2.87 & 0.00 & 1.00  \\ 
    &   1.40 &   2 & -82.85 & 34.68 & 0.75 & 0.00 & 1.00  \\ 
    &   1.40 &   2 & -82.85 & 34.68 & 1.00 & 0.00 & 2.00 \\ 
    &   1.50 &   1 & -82.85 & 34.68 & 3.19 & 0.00 & 1.00  \\ 
    &   1.50 &   2 & -82.85 & 34.68 & 1.24 & 0.00 & 1.00  \\ 
    \hline
   FALSE  & 2.00 &   1 & -82.85 & 34.68 & 4.15 & 0.00 & 1.00  \\
    &   2.00 &   2 & -82.85 & 34.68 & 2.26 & 0.00 & 1.00  \\ 
    &   2.00 &   3 & -82.04 & 34.16 & 0.00 & 72.32 & 1.00  \\ 
    &   2.10 &   1 & -82.85 & 34.68 & 4.31 & 0.00 & 1.00 \\ 
    &   2.10 &   2 & -82.85 & 34.68 & 2.48 & 0.00 & 1.00  \\ 
    &   2.10 &   3 & -82.81 & 34.30 & 0.26 & 71.20 & 1.00  \\ 
    &   2.20 &   1 & -82.85 & 34.68 & 4.57 & 0.00 & 1.00  \\ 
    &   2.20 &   2 & -82.85 & 34.68 & 2.62 & 0.00 & 1.00  \\ 
\noalign{\smallskip}\hline\noalign{\smallskip}
\end{tabular}
\end{table}
\subsection{Attack Detection Framework}
\label{sctdet}
Figure~\ref{fig_flow} depicts the approach of attack detection using EM and CUSUM. In order to implement change point detection, first step is to identify the processing time window in which information need to track, and how to convert such information into sequence to detect shifts due to malicious attacks and/or benign system malfunctions. In DoS or flooding attacks, vehicles are expected to send more messages than the designed frequency parameter (e.g., MVS). Therefore, tracking messages per vehicle and estimating MVS can be used as indicators for detection. For impersonation attack, multiple messages in unit time interval ($0.1$ s) are sent, and by monitoring MVT, this type of attack is detected. Lastly, false information can be defined as any type of irregularity in the collected messages, such as high or low vehicle speed compared to rest of the traffic (inherent) at a roadway segment or an unrealistic gap between any two adjacent vehicles within a certain time frame. CUSUM monitors deviations from process mean and identifies violations. On the other hand, EM calculates conditional probabilities of $P(DoS attack|MVS)>0.001$, where $P(impersonation|MVT)$ and $P(attack state|distance)$ are given. Simply, if the likelihood at any time is $>0.001$, then an attack is detected.  
\begin{figure}[H]
\centering
  \includegraphics[width=0.75\linewidth]{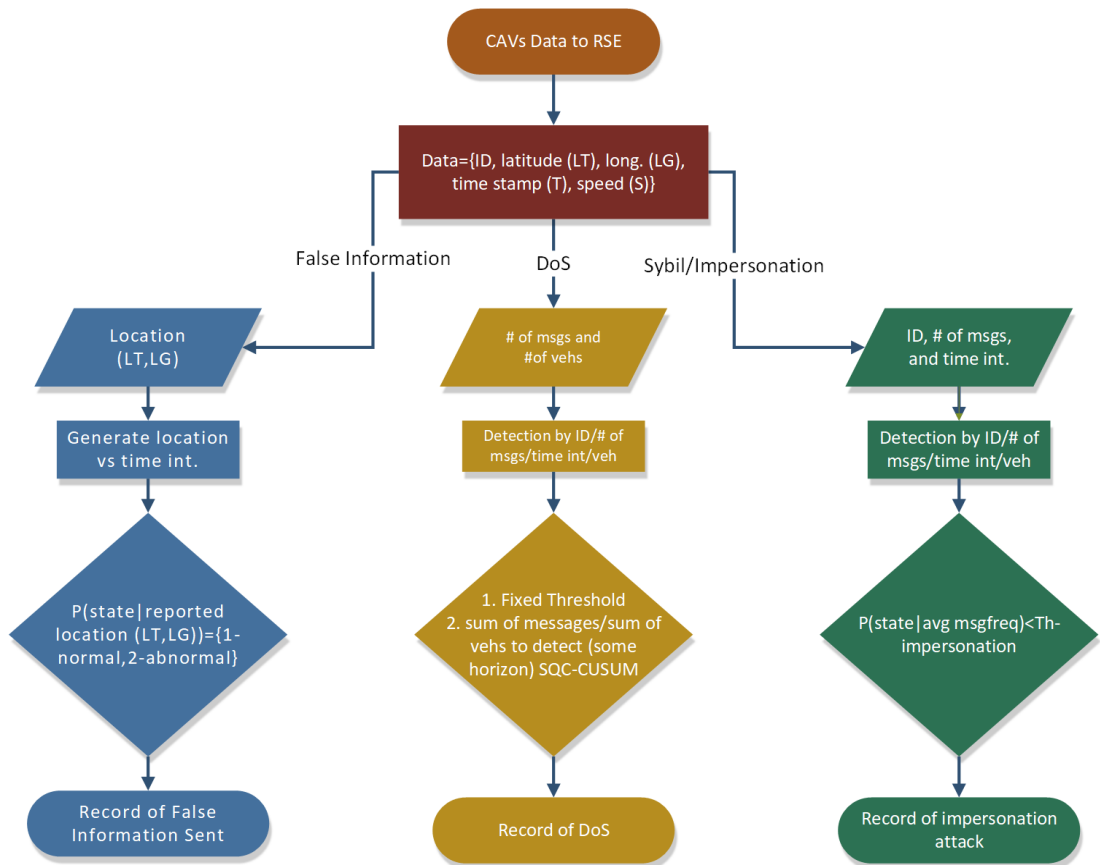}
\caption{Attack detection approach in V2I connected vehicle environment.}
  \label{fig_flow}
\end{figure}%

In Table~\ref{attacks_tab}, position of a vehicle column is calculated in meters (m) from two consecutive latitude and longitude values. MST and MSV measures are deduced from time and ID columns for every time interval of $0.1$ s and time series are generated for statistical detection. It should also be noted that for the DoS attack vehicle number $6$ is not sending speed and location correctly. Attack detection using the change of speed and distance would be trivial. Attacker would also replicate reasonable values. So, detection is carried out using message frequency in MSV. From the table, EM's $P(attack|observation)$ is denoted as $P(D|Y_t)>0.001$ resulting as detection (D) otherwise no detection (ND). Similarly for CUSUM, $(C^+,C^-)$ values are given. Based on these values, when $(C^+ or C^-)>5\sigma$ a detection is observed, otherwise ND is issued. Persistent attacks are easily detected by CUSUM and EM. CUSUM continues to detect normal observations as attacks (i.e.,an out-of-control process) and generates false positive errors. This can be fixed in CUSUM with a slight revision in $C^-$ values mimicking one-sided control. 
\begin{table}[H]
\centering
\caption{Examples of attack data on RSE and detection by EM and CUSUM}
\label{attacks_tab}
\scalebox{0.8}{
\begin{tabular}{lrrrrrrrlrl}
 Type & Freq. & TS(s) & ID & Spd(m/s) & Pos.(m) & Msgs. & $P(D|Y_t)$ & EM & $(C^+,C^-)$& CUS\\ 
  \hline
DoS & 127 & 5.00 &   1 & 9.73 & 0.08 & 10.00 & 0.00 & ND &(0,0)&ND\\ 
 &128 & 5.00 &   2  & 8.08 & 0.52 & 10.00 & 0.00 & ND &(0,0)&ND\\ 
    &129 & 5.00 &   3 &  5.97 & 0.74 & 10.00 & 0.00 & ND& (0,0)&ND\\ 
     &130 & 5.00 &   5 &  2.14 & 0.40 & 10.00 & 0.00 & ND& (0,0)&ND\\ 
    &131 & 5.10 &   1 & 9.94 & 0.08 & 10.00 & 0.00 & ND &(0,0)&ND\\ 
     &132 & 5.10 &   2 &  8.22 & 0.52 & 10.00 & 0.00 & ND& (0,0)&ND\\ 
     &133 & 5.10 &   3 & 6.21 & 0.74 & 10.00 & 0.00 & ND& (0,0)&ND\\ 
     &134 & 5.10 &   5 &  2.32 & 0.14 & 10.00 & 0.00 & ND& (0,0)&ND\\ 
     &135 & 5.10 &   6 &  0.00 & 0.00 & 10.12 & 0.02 & D& (0.12,0)&D\\ 
     &136 & 5.10 &   6 &  0.00 & 0.00 & 10.23 & 0.05 & D& (0.12,0)&D\\ 
     &137 & 5.10 &   6 &  0.00 & 0.00 & 10.35 & 0.09 & D& (0.29,0)&D\\ 
    \hline
   IMP &  13 & 1.20 &   1  & 2.44 & 0.00 & 1.00 & 0.00 & ND &(0,0) &ND\\
    &  14 & 1.20 &   2 &  0.26 & 0.00 & 1.00 & 0.00 & ND & (0,0)&ND\\ 
    &  15 & 1.30 &   1 &  2.65 & 0.00 & 1.00 & 0.00 & ND & (0,0)&ND\\ 
    &  16 & 1.30 &   2 &  0.51 & 0.00 & 1.00 & 0.00 & ND & (0,0)&ND\\ 
    &  17 & 1.40 &   1 &  2.87 & 0.00 & 1.00 & 0.00 & ND & (0,0)&ND\\ 
    &  18 & 1.40 &   2 &  0.75 & 0.00 & 1.00 & 0.00 & ND & (0,0)&ND\\ 
    &  19 & 1.40 &   2 &  1.00 & 0.00 & 2.00 & 0.67 & D &0.99,0 &D\\ 
    &  20 & 1.50 &   1 &  3.19 & 0.00 & 1.00 & 0.00 & ND & 0,0.99&D\\ 
    &  21 & 1.50 &   2 &  1.24 & 0.00 & 1.00 & 0.00 & ND & 0.99,0&D\\ 
    \hline
   FLS &  31 & 2.00 &   1  & 4.15 & 0.00 & 1.00 & 0.00 & ND & (0,0) &ND \\
    &  32 & 2.00 &   2 &  2.26 & 0.00 & 1.00 & 0.00 & ND & (0,0)&ND\\ 
    &  33 & 2.00 &   3  & 0.00 & 72.32 & 1.00 & 1.00 & D & (72.2,0)&D\\ 
    &  34 & 2.10 &   1  & 4.31 & 0.00 & 1.00 & 0.00 & ND & (0,72.3)&D\\ 
    &  35 & 2.10 &   2 &  2.48 & 0.00 & 1.00 & 0.00 & ND & (72.2,0)&D\\ 
    &  36 & 2.10 &   3 &  0.26 & 71.20 & 1.00 & 1.00 & D & (34.9,0)&D\\ 
    &  37 & 2.20 &   1 & 4.57 & 0.00 & 1.00 & 0.00 & ND & (0,11.7)&D\\ 
    &  38 & 2.20 &   2 & 2.62 & 0.00 & 1.00 & 0.00 & ND & (5.8,0)&D\\ 
    &  39 & 2.20 &   3 &  0.52 & 49.76 & 1.00 & 1.00 & D & (48.2,0)&D\\ 
    &  40 & 2.30 &   1 & 4.81 & 0.00 & 1.00 & 0.00 & ND & (0,9.7)&D\\ 
    &  41 & 2.30 &   2 &  2.83 & 0.00 & 1.00 & 0.00 & ND & (3.2,0)&D\\ 
    &  42 & 2.30 &   3 &  0.75 & 34.29 & 1.00 & 1.00 & D & (33.6,0)&D\\ 
    &  43 & 2.40 &   1 &  4.95 & 0.00 & 1.00 & 0.00 & ND & (0,4.9)&D\\ 
\noalign{\smallskip}\hline\noalign{\smallskip}
\end{tabular}
}
\end{table}

\section{Conclusions}
\label{sctconc}
In this study, we investigated the efficacy of two statistical change point models, EM and CUSUM, for real-time V2I cyberattack detection in a CV Environment. We evaluated these two models for three different type of cyberattacks, DoS, impersonation, and false information, using BSMs. Instead of tracking data values such as message frequency, speed, and distance individually for each time interval and vehicle ID, aggregate measures are deduced from BSMs to be used in effective real-time detection. Based on the numerical analysis, we found that: 
\begin{enumerate}[label=(\roman*)]
\item Given proper initialization, i.e., mean and variance of normal and abnormal cases, and enough computational power, both algorithms can detect all three attack types accurately.
\item When attack detection time window is critical such as safety applications, for EM, it is greater than  $<0.1$ second, whereas, the time window for CUSUM is below $0.1$ second. 
\item When multiple states could be observed for an attack or to classify different impacts, as well as any changes in the normal roadside unit communication frequencies, EM algorithm would be able to provide conditional probabilities for multiple states.
\end{enumerate}
Results from numerical analysis also revealed that both EM and CUSUM could detect these cyberattacks with accuracy of at least 98\% and 98\%, respectively. Both models can be applied for real-time cyber attack detection with a one-second interval. Possible improvements to this research can be listed as: (1) further research is needed to investigate factors affecting the optimal selection of such parameters with multiple data sets; (2) comprehensive attack modeling can be developed.

\bibliographystyle{trb}
\bibliography{ms}

\end{nolinenumbers}
\end{document}